\documentclass[prl,twocolumn,floatfix,amsmath,nofootinbib,amssymb,floatfix]{revtex4}
\usepackage{graphicx,color,dcolumn,booktabs,bm}
\usepackage{longtable,lscape}
\usepackage{txfonts}
\usepackage{overpic}
\usepackage{amssymb}
\usepackage{indentfirst}
\usepackage{feynmf}   %{feynmp}
\usepackage{slashed}  %for Feynman symbols
\usepackage{cases}
\usepackage{color}
\usepackage{multirow}
\usepackage{epstopdf}
\usepackage{enumerate}
\usepackage{graphicx,color,dcolumn,booktabs,bm}
\usepackage[colorlinks, citecolor=blue,anchorcolor=red,menucolor=red, linkcolor=red,filecolor=red,urlcolor=blue,frenchlinks=red]{hyperref}

\begin{document}
%\begin{CJK}{GBK}{}

\title{Producing fully-charm structures in the $J/\psi$-pair invariant mass spectrum}
\author{Jun-Zhang Wang$^{1,2}$}\email{wangjzh2012@lzu.edu.cn}
\author{Dian-Yong Chen$^3$}\email{chendy@seu.edu.cn}
\author{Xiang Liu$^{1,2}$\footnote{Corresponding author}}\email{xiangliu@lzu.edu.cn}
\author{Takayuki Matsuki$^{4}$}\email{matsuki@tokyo-kasei.ac.jp}
\affiliation{$^1$School of Physical Science and Technology, Lanzhou University, Lanzhou 730000, China\\
$^2$Research Center for Hadron and CSR Physics, Lanzhou University $\&$ Institute of Modern Physics of CAS, Lanzhou 730000, China\\
$^3$School of Physics, Southeast University, Nanjing 210094, China\\
$^4$Tokyo Kasei University, 1-18-1 Kaga, Itabashi, Tokyo 173-8602, Japan}

\date{\today}

\begin{abstract}
Very recently, the LHCb Collaboration reported the observation of several enhancements in the invariant mass spectrum of a $J/\psi$ pair between 6.2 and 7.4 GeV. In this work, we propose the dynamical mechanism to mimic the experimental data of a di-$J/\psi$ mass spectrum given by LHCb, which is based on the reactions, where all the possible combinations of a double charmonium directly produced by a  proton-proton collision are transferred into a final state $J/\psi J/\psi$. We find that the LHCb experimental data can be well reproduced. We further extend our framework to study a di-$\Upsilon(1S)$ system, and give the line shape of a differential cross section of a partner process in a $b\bar{b}$ system on the invariant mass of $\Upsilon(1S)\Upsilon(1S)$, which shows that there should exist possible enhancements near $m_{\Upsilon(1S)\Upsilon(1S)}=$19.0, 19.3, 19.7 GeV in the $\Upsilon(1S)$-pair invariant mass spectrum. These predictions can be tested in LHCb and CMS, which can be as a potential research issue in near future.
\end{abstract}

\maketitle

%\section{Introduction}\label{sec1}

{\it Introduction.---}Since 2003, benefitted from the accumulation of more and more experimental data, searches for exotic multiquark matter have become a hot issue in hadron physics and attracted more and more interests from theorists and experimentalists.  The important progress for this topic is the observations of a number of charmoniumlike $XYZ$ states and the hidden-charm pentaquark $P_c$ states in the relevant high energy experiments (see review articles \cite{Chen:2016qju,Liu:2019zoy,Guo:2017jvc,Olsen:2017bmm,Brambilla:2019esw} for more details). The study of these novel phenomena have indeed enlarged our knowledge for the nonperturbative quantum chromodynamics (QCD).

Very recently, the LHCb Collaboration measured the $J/\psi$-pair mass spectrum by using proton-proton data at center of mass energies of 7, 8, and 13 TeV, where a narrow structure around 6.9 GeV was observed with a significant signal of more than standard deviation of 5.1$\sigma$ \cite{Aaij:2020fnh}. Besides, there exist an obvious broad structure ranging from the threshold of di-$J/\psi$ to 6.8 GeV and an underlying peak near 7.3 GeV. Assuming $X(6900)$ and that the other two peaks are produced by hadronic resonances, it can be easily concluded that they are composed of four charm quarks  $cc\bar{c}\bar{c}$, which is deduced from the measured final states of a double $J/\psi$. Thus, this special quark configuration determines that these newly observed structures are not likely explained as  exotic molecular state, hybrid, etc., and interpretation of  a compact fully-charm tetraquark state is the most mainstream. 

Indeed, several theoretical articles have recently focused on the researches for fully-heavy tetraquark states and discussed the possible assignment of $X(6900)$ in the corresponding spectroscopy \cite{Chen:2020xwe,Jin:2020jfc,Wang:2020ols,Lu:2020cns,Yang:2020rih,Deng:2020iqw,Chen:2020lgj,Albuquerque:2020hio,Sonnenschein:2020nwn,Giron:2020wpx,Richard:2020hdw,Becchi:2020uvq,liu:2020eha}. In fact, as early as in 1981, the exotic hadrons composed of $cc\bar{c}\bar{c}$ have been systematically studied by Chao for the first time \cite{Chao:1980dv}, which has predicted that fully-charm tetraquark states are all above the threshold of strong decay into two charmonia. Later, a similar conclusion was obtained in Refs. \cite{Ader:1981db,SilvestreBrac:1993ss}. On the other hand, in Ref. \cite{Heller:1985cb}, the authors used an MIT bag model with the Born-Oppenheimer approximation for heavy quarks and found that there exists a stable fully-charm tetraquark against breakup into $c\bar{c}$ pairs. In the following decades, the mass spectra for fully-heavy quark states had been still hotly debated in various model schemes. For instance, the quark potential model \cite{Lloyd:2003yc,Barnea:2006sd,Wu:2016vtq,Debastiani:2017msn,Wang:2019rdo,Liu:2019zuc,Bedolla:2019zwg}, the QCD sum rule \cite{Chen:2016jxd,Chen:2018cqz,Wang:2018poa}, the non-relativistic effective field theory \cite{Anwar:2017toa}, the covariant Bethe-Salpeter equations \cite{Heupel:2012ua}, and other phenomenological methods \cite{Berezhnoy:2011xy,Berezhnoy:2011xn,Karliner:2016zzc,Karliner:2017qhf}. Anyway, all the studies mentioned above supported the existence of exotic fully-heavy quark hadrons although this conclusion was also questioned in some theoretical papers \cite{Richard:2017vry,Czarnecki:2017vco}.

Actually, the Large Hadron Collider (LHC) is an ideal experimental platform to search for the fully-charm structures with the $cc\bar{c}\bar{c}$ configuration, where its production can be achieved by hadronization of four charm quarks  in the single parton scattering (SPS) \cite{Sun:2014gca,Likhoded:2016zmk,Baranov:2011zz,Lansberg:2013qka,Lansberg:2014swa,Lansberg:2015lva,Shao:2012iz,Shao:2015vga} process of $gg\to c\bar{c}c\bar{c} + X$. Then, it can dominantly decay into a pair of charmonium states like $\eta_c\eta_c$,  $J/\psi J/\psi$, etc., through the so-called fall-apart decay mechanism \cite{Chen:2020xwe,Zhu:2005hp,Giacosa:2006rg}. Among the allowed final states, the $J/\psi J/\psi$ final state is the most promising candidate to search for fully-charm structures above the production threshold since a $J/\psi$ particle can be effectively reconstructed by a  $\mu^+\mu^-$ pair via a muon detector. 
On the other hand, the more double $J/\psi$ events can be directly produced by the single parton scattering (SPS) \cite{Sun:2014gca,Likhoded:2016zmk,Baranov:2011zz,Lansberg:2013qka,Lansberg:2014swa,Lansberg:2015lva,Shao:2012iz,Shao:2015vga} and the double parton scattering (DPS) \cite{Calucci:1997ii,Calucci:1999yz,DelFabbro:2000ds} processes in high energy proton-proton collisions, which usually correspond to a continuum contribution to the invariant mass spectrum of  $J/\psi J/\psi$. Thus, it seems that the several structures in the mass spectrum of di-$J/\psi $ observed by LHCb are indeed good candidates of fully-charm tetraquark states. The origin of resonance peak phenomena in hadron physics are, however, usually complicated more than our common understanding. Hence, the nature of the new structure $X(6900)$ observed by LHCb has to be judged carefully.   

Although the fully-charm tetraquark seems to be a dominant explanation for the newly observed structure in the $J/\psi$-pair invariant mass spectrum \cite{Aaij:2020fnh}. It should be emphasized that in the vicinity of $X(6900)$, there are abundant thresholds of charmonium pair, such as $\eta_c(1S)\chi_{c1}(1P)$, $\chi_{c0}(1P)\chi_{c1}(1P)$ and $\chi_{c0}(1P)\chi_{c1}^{\prime}(2P)$. These charmonium pair can transit into di-$J/\psi$ by rescattering and the lineshapes of the rescattering channels will appear some structures due to threshold cusps. Such kind of threshold cusps will provide non-trivial contributions, which can not be described by simple smooth functions. In this letter, we propose a dynamical rescattering mechanism to understand the newly observed $X(6900)$ and other two structures in the di-$J/\psi$ invariant spectrum. {The present investigation not only provides a novel interpretation to $X(6900)$, but more importantly, reminds us to carefully check the non-resonance contributions before treating the structure as a genuine resonance.}

\begin{figure}[htb]
	\includegraphics[width=8cm,keepaspectratio]{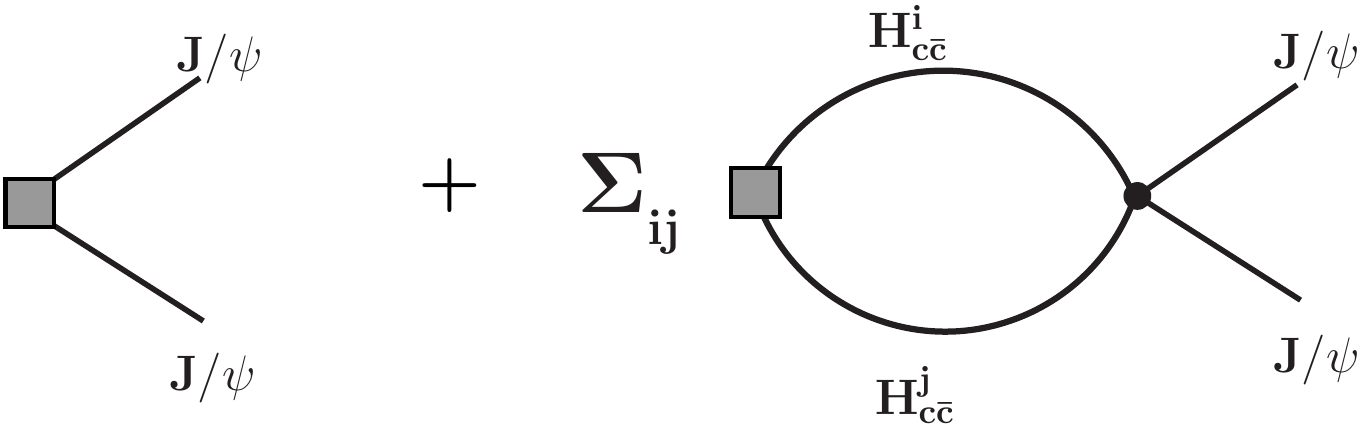}
	\caption{ The schematic diagrams for the production mechanism of a double charmonium $J/\psi J/\psi$, where $H^{i}_{c\bar{c}}$ stands for allowed intermediate charmonium states, such as $\eta_c$, $J/\psi$, $\chi_{cJ}$ with $J=0,1,2$, etc. Here, the gray rectangle corresponds to direct production of a double charmonium in hadron collisions. \label{fig:feynman}  }
\end{figure}

{\it Dynamical mechanism}.---Various combinations of double charmonia can be directly produced by both the SPS and DPS processes. Then, the double charmonium allowed by quantum numbers from direct production can be transferred into final states of $J/\psi J/\psi$. The relevant schematic diagrams are shown in Fig. \ref{fig:feynman}, where the interaction among intermediate charmonium pairs and $J/\psi $ pairs are absorbed into a vertex.

Starting from an $S$-wave interaction between intermediate charmonium pairs, the production amplitude of double $J/\psi$ by rescattering mechanism in Fig. \ref{fig:feynman} becomes the one proportional to the scalar two-point loop integral, whose analytical form can be given by, in the rest frame of di-$J/\psi $,
\begin{eqnarray}
&&L_{ij}(m_{J/\psi J/\psi})=\int \frac{dq^4}{(2\pi)^4} \frac{e^{-(2\vec{q}~)^{2}/\alpha^2}}{(q^2-m_{i}^2+i\epsilon)((P-q)^2-m_{j}^2+i\epsilon)} \nonumber \\
&&~=\frac{i}{4m_{i}m_{j}}
\left\{\frac{-\mu\alpha}{\sqrt{2}(2\pi)^{3/2}}+\frac{\mu\sqrt{2\mu m_0}\left(\textrm{erfi}\left[\frac{\sqrt{8\mu m_0}}{\alpha}\right]-i\right)}{2\pi/e^{-\frac{8\mu m_0}{\alpha^2}}} 
\right\}, \label{eq:1}
\end{eqnarray}
where $m_i$($m_j$) is the mass of an intermediate charmonium state $H_{c\bar{c}}^i$($H_{c\bar{c}}^j$) marked in Fig. \ref{fig:feynman} and $\mu=(m_im_j)/(m_i+m_j)$ and $m_0=m_{J/\psi J/\psi}-m_{i}-m_{j}$.  $P=(m_{J/\psi J/\psi},0,0,0)$ stands for the four-momentum of a double $J/\psi $ system and  the erfi$(x)$ represents  the imaginary error function. Here, an exponential form factor $e^{-(2\vec{q}~)^{2}/\alpha^2}$ is introduced to avoid the ultraviolet divergence of scalar two-point loop integral, and $\alpha$ is a cutoff parameter.

For the one-loop rescattering processes in Fig. \ref{fig:feynman}, there exists a square root branch point, $\sqrt{m_{J/\psi J/\psi}-m_{i}-m_{j}}$, where an integral singularity at the threshold of $m_{i}+m_{j}$ appears at the on-shell of two intermediate charmonium states. The threshold singularity causes a cusp exactly at the corresponding threshold  in the invariant mass distribution of $m_{J/\psi J/\psi}$. Here, according to the sequence of  magnitude of thresholds and criterion of quantum number conservation, $H_{c\bar{c}}^iH_{c\bar{c}}^j$= $J/\psi J/\psi$, $\eta_c\chi_{cJ}$, $J/\psi h_c$, $\chi_{cJ}\chi_{cJ}$, and $\chi_{c0} \chi_{c1}^{\prime}$ with $J=0,1,2$ are selected. In this work, without any special emphasis, $\eta_c$, $h_c$, $\chi_{cJ}$, and $\chi_{c1}^{\prime}$ refer to $\eta_c(1S)$, $h_c(1P)$, $\chi_{cJ}(1P)$, and $\chi_{c1}^{\prime}(2P)$, respectively, and $\chi_{c1}^{\prime}(2P)=X(3872)$ \cite{Kalashnikova:2005ui,Zhang:2009bv,Kalashnikova:2009gt,Li:2009ad,Coito:2010if}. This selection covers the energy region from 6.194 to 7.400 GeV, in which there exist several relatively clear peaks measured by LHCb \cite{Aaij:2020fnh}. 
%From Fig. \ref{fig:Fitexp},  we can see that the rescattering processes indeed produce the obvious threshold cusps and its visibility depends on the cutoff parameter $\alpha$. 
It is worth emphasizing that the direct production rates of $\eta_c$, $X(3872)$, and $P$-wave charmonium states $\chi_{cJ}$ with $J=0,1,2$ in high energy proton-proton collision have been proved to be comparable with that of $J/\psi$ particle by both experiments \cite{Aaij:2014bga,Aaij:2011sn} and theoretical calculations from nonrelativistic QCD(NRQCD) \cite{Bodwin:1994jh,Ma:2014mri,Li:2011yc,Butenschoen:2014dra,Han:2014jya,Bodwin:2015iua,Ma:2010vd,Artoisenet:2009wk,Butenschoen:2013pxa}.  Combined with an $S$-wave coupling on the production vertex of their combinations, it is fully possible that the resonance-like shapes from the rescattering processes are responsible for the recent LHCb data without introducing any hadronic resonances of configuration of $cc\bar{c}\bar{c}$. In the following, in order to verify the above idea, we will focus on the experimental line shape of an invariant mass spectrum of di-$J/\psi$ by LHCb.

\iffalse
\begin{figure}[b]
	\includegraphics[width=8.7cm,keepaspectratio]{Lineshape.pdf}
	\caption{ The dependence of scalar two-point loop integral $\mid L \mid^2$ on the invariant mass of $m_{J/\psi J/\psi}$, where a cutoff $\alpha=3$ is taken and the allowed combinations of a double charmonium by the $C$ parity include $J/\psi J/\psi$, $\eta_c\chi_{cJ}$, $J/\psi h_c$, $\chi_{cJ}\chi_{cJ}$, and $\chi_{c0} \chi_{c1}^{\prime}$ with $J=0,1,2$. In the last diagram, the comparisons of line shapes of $\mid L \mid^2$ among different values of $\alpha$ are presented by taking $\chi_{c0}\chi_{c1}$ as an example. Here, the maximums are normalized to one. \label{fig:lineshape}  }
\end{figure}
\fi

The direct production of a double charmonium from SPS and DPS processes in high energy proton-proton collisions usually behaves like a continuous distribution in its invariant mass spectrum, whose complete estimation is very complicated and parameter-dependent in theoretical approaches \cite{Sun:2014gca,Likhoded:2016zmk,Baranov:2011zz,Lansberg:2013qka,Lansberg:2014swa,Lansberg:2015lva,He:2019qqr,He:2015qya,Lansberg:2020rft,Lansberg:2019fgm,Li:2009ug}.  So in this work, referring to the treatment of experimental analysis of LHCb \cite{Aaij:2020fnh}, the invariant mass distribution of direct production of a double charmonium $H_{c\bar{c}}^i H_{c\bar{c}}^j$ with an $S$-wave can be parameterized as
\begin{eqnarray}
\mathcal{A}^2_{direct}=g^2_{direct}~e^{c_0m_{ij}}~\frac{1}{8\pi}\frac{\sqrt{\lambda(m_{ij}^2,m_{i}^2,m_{j}^2)}}{m_{ij}^2},
\end{eqnarray}
where $\lambda(x^2,y^2,z^2)=x^2+y^2+z^2-2xy-2xz-2yz$ is the K$\ddot{\textrm{a}}$llen function, and $m_{ij}$ is the corresponding invariant mass. For the rescattering processes with two types of intermediate charmonium pairs $H_{c\bar{c}}^i H_{c\bar{c}}^j=J/\psi J/\psi, \chi_{cJ}\chi_{cJ}, \chi_{c0}\chi_{c1}^{\prime}$ and $H_{c\bar{c}}^i H_{c\bar{c}}^j=\eta_c\chi_{cJ}, J/\psi h_c$ in Fig. \ref{fig:feynman}, the line shapes on the invariant mass spectrum of $m_{J/\psi J/\psi}$ are given by
\begin{eqnarray}
\mathcal{A}^2_{ij}(m_{J/\psi J/\psi})=g_{ij}^2 L_{ij}^2(m_{J/\psi J/\psi})\frac{e^{c_0m_{J/\psi J/\psi}}p_{J/\psi}}{m_{J/\psi J/\psi}}
\end{eqnarray}
and
\begin{eqnarray}
\mathcal{A}^{\prime2}_{ij}(m_{J/\psi J/\psi})=g_{ij}^2 L_{ij}^2(m_{J/\psi J/\psi})\frac{e^{c_0^{\prime}m_{J/\psi J/\psi}}p_{J/\psi}^3}{m_{J/\psi J/\psi}},
\end{eqnarray}
respectively, where $p_{J/\psi}$ is the momentum of a final state $J/\psi$, and $e^{c_0^{(\prime)}m_{J/\psi J/\psi}}$ is from the direct production vertex. It is worth noting that the system parity of the above two types of rescattering processes are $P=+1$ and $P=-1$, respectively. Thus, the total line shape for the invariant mass distribution of producing a double $J/\psi$ in high energy proton-proton collisions can be written as
\begin{eqnarray}
\mathcal{A}^2&=\mid \mathcal{A}_{direct}(m_{J/\psi J/\psi})+\sum_{mn}e^{i\phi^{mn}}\mathcal{A}_{mn}(m_{J/\psi J/\psi}) \mid^2  \nonumber \\
&+\mid \mathcal{A}_{direct}^{\prime}(m_{J/\psi J/\psi})+\sum_{mn}e^{i\phi^{mn}}\mathcal{A}^{\prime}_{mn}(m_{J/\psi J/\psi}) \mid^2 
\end{eqnarray}
where $\phi^{mn}$ is the phase between direct contribution and the corresponding rescattering process. 
Here, we introduce a background term $\mathcal{A}_{direct}^{\prime}=(\frac{g^{\prime2}_{direct}\lambda(m_{J/\psi J/\psi}^2,m_{J/\psi}^2,m_{J/\psi}^2)^{\frac{1}{2}}e^{c_0^{\prime}m_{J/\psi J/\psi}}p_{J/\psi}^2}{8\pi m_{J/\psi J/\psi}^2})^{\frac{1}{2}}$  for the rescattering amplitude $\mathcal{A}^{\prime}_{mn}$ with $P=-1$, which corresponds to a direct production of the $P$-wave double $J/\psi$.

{\it Lineshape of $J/\psi$ pair invariant mass spectrum.}---With the above preparations, we can directly study the LHCb data based on our proposed dynamical mechanism. Firstly, it can be seen that there exist eleven predicted threshold cusps at the di-$J/\psi$ energy region from 6.194 to 7.400 GeV, which far exceed the number of visible obvious peak structures observed by LHCb and will bring some difficulties in our theoretical analysis. Fortunately, due to the property of an approximate mass degeneracy among  three $P$-wave charmonium $\chi_{cJ}$ states, we find that the threshold cusps are actually and mainly concentrated in five energy positions, i.e., ($6.45\sim6.58$), $6.64$, ($6.87\sim7.00$), ($7.03\sim7.13$) and $7.32$ GeV. With present experimental precision, it must be difficult to distinguish the individual signals from these close peaks, where their contributions may overlap and so behave like one peak structure. Thus, in the realistic analysis of experimental LHCb data, we only consider the rescattering processes from $J/\psi J/\psi$, $\eta_c\chi_{c1}$, $J/\psi h_c$, $\chi_{c0}\chi_{c1}$, and $\chi_{c0} \chi_{c1}^{\prime}$, where $\eta_c\chi_{c1}$ and $\chi_{c0}\chi_{c1}$ are representative channels which approximately contain all of the contributions from $\eta_c\chi_{cJ}$ and $\chi_{c0}\chi_{cJ}$ with $J=0,1,2$, respectively. To further reducing the fitting parameters, we notice that there is no evident structure in the energy region between 7.03 and 7.13 GeV of the LHCb data with present precision, thus we exclude the contributions from  $\chi_{c1}\chi_{c1}$, $\chi_{c1}\chi_{c2}$, and $\chi_{c2}\chi_{c2}$.

\begin{figure}[t]
\includegraphics[width=8.7cm,keepaspectratio]{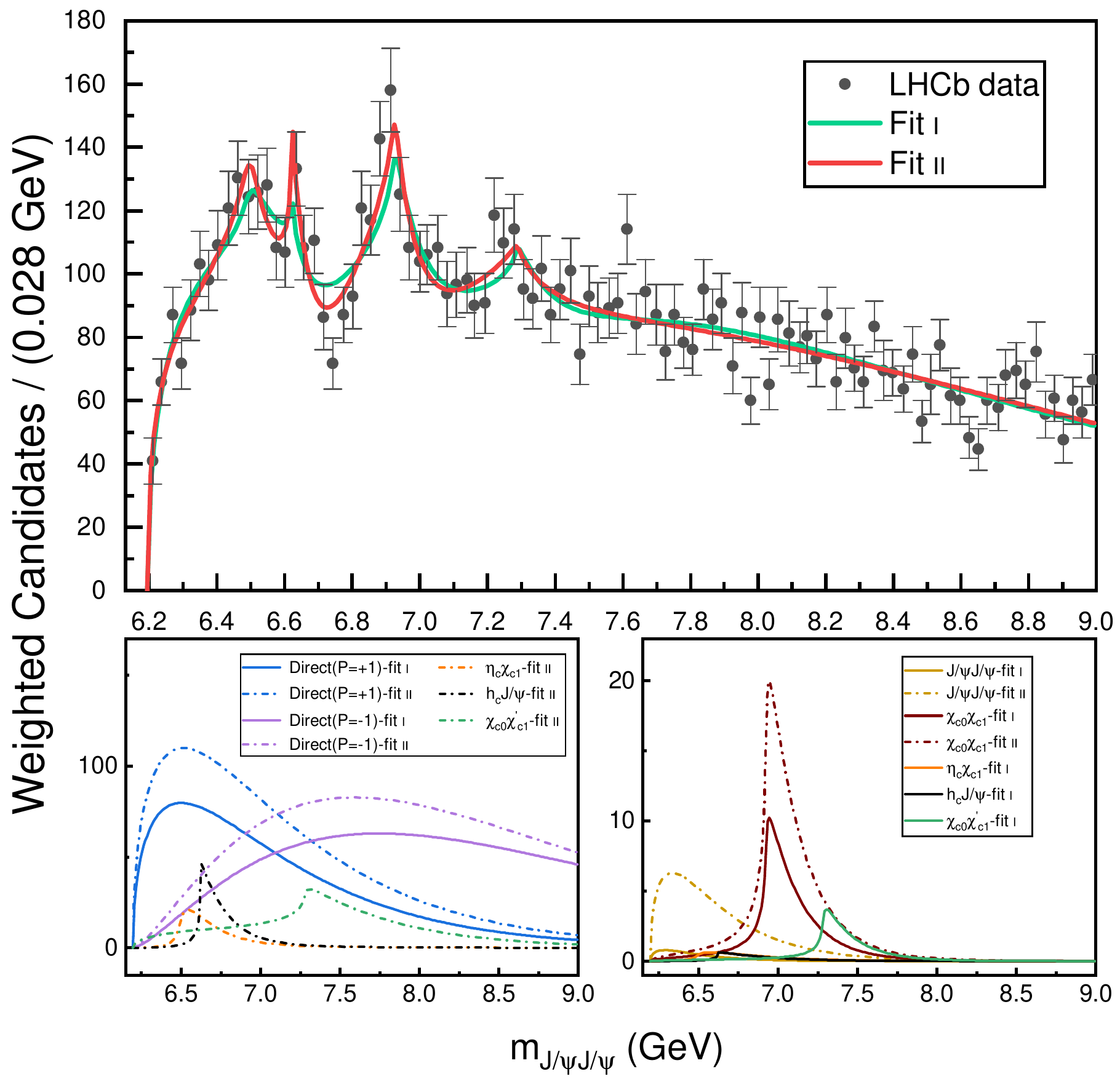}
	\caption{ The fit of our line shape to the LHCb data \cite{Aaij:2020fnh} based on a dynamical rescattering mechanism. Here, two fitting schemes of fit I and fit II are introduced, which correspond to the same and independent cutoffs $\alpha$ for different combinations of intermediate charmonium states selected in the fitting procedure, respectively.  \label{fig:Fitexp}  }
\end{figure}

\begin{table}[h]
  	\caption{The parameters for reproducing the line shape of LHCb data in two fitting schemes of fit I and fit II.}
  	\setlength{\tabcolsep}{3.6mm}{
  	\begin{tabular}{cccccccc}
			\toprule[1.0pt]
     Parameters &  Fit I  &  Fit II  \\
			\toprule[1.0pt]
          $c_0$ ~($\textrm{GeV}^{-1}$) & $-1.52\pm0.02$  & $-1.45\pm0.01$   \\ 
          $c_0^{\prime}$ ~($\textrm{GeV}^{-1}$) & $-0.946\pm0.058$  & $-1.05\pm0.01$   \\ 
          $\mid g_{direct}^{\prime}/g_{direct} \mid$  & $0.0767\pm0.0204$  & $0.137\pm0.042$    \\ 
            $\mid g_{J/\psi J/\psi}/g_{direct} \mid$  & $8.53\pm3.64$  & $14.0\pm1.4$    \\ 
            $\mid g_{\eta_c\chi_{c1}}/g_{direct}^{\prime} \mid$  & $91.6\pm75.4$  & $112\pm28$    \\ 
             $\mid g_{J/\psi h_c}/g_{direct}^{\prime} \mid$  & $69.7\pm16.1$ & $109\pm8$    \\
             $\mid g_{\chi_{c0}\chi_{c1}}/g_{direct} \mid$  & $33.3\pm8.2$  & $38.5\pm7.6$    \\ 
              $\mid g_{\chi_{c0} \chi_{c1}^{\prime}}/g_{direct} \mid$  & $25.8\pm10.6$  & $19.0\pm4.3$    \\ 
              $\phi_{J/\psi J/\psi}$ ~(rad) & $1.53\pm0.51$  & $3.16\pm0.19$    \\ 
              $\phi_{\eta_c\chi_{c1}}$ ~(rad) & $2.69\pm0.20$  & $2.80\pm0.15$    \\             
              $\phi_{J/\psi h_c}$ ~(rad)  & $4.40\pm0.33$  & $2.95\pm0.24$    \\ 
               $\phi_{\chi_{c0}\chi_{c1}}$ ~(rad) & $2.14\pm0.18$  & $2.89\pm0.20$    \\ 
               $\phi_{\chi_{c0} \chi_{c1}^{\prime}}$ ~(rad) & $2.00\pm0.33$  & $3.23\pm0.20$    \\ 
               $\alpha_{J/\psi J/\psi} $ ~(GeV)  & $1.71\pm0.01$  & $2.30\pm0.21$    \\                
               $\alpha_{\eta_c\chi_{c1}} $ ~(GeV) & $1.71\pm0.01$  & $1.20\pm0.21$    \\                                 
                  $\alpha_{J/\psi h_c} $ ~(GeV) & $1.71\pm0.01$  & $1.20\pm0.03$    \\                                  
                    $\alpha_{\chi_{c0}\chi_{c1}} $ ~(GeV) & $1.71\pm0.01$  & $1.73\pm0.26$    \\                                        
                      $\alpha_{\chi_{c0} \chi_{c1}^{\prime}} $ ~(GeV) & $1.71\pm0.01$  & $5.20\pm0.05$    \\   
                     \bottomrule[0.6pt]     
                      $\chi^2/d.o.f$ & 1.41  & 1.25    \\  
			\bottomrule[1.0pt]
		\end{tabular}\label{table:parameter}}
  \end{table}

In Fig. \ref{fig:Fitexp}, our theoretical fit to the experimental line shape of invariant mass spectrum vs. $m_{J/\psi J/\psi}$ in the high energy proton-proton collisions is given based on the dynamical rescattering mechanism in Fig. \ref{fig:feynman}. The relevant fitting parameters are listed in Table \ref{table:parameter}. Here, it is worth emphasizing that widths of some of charmonium states such as $\eta_c$ are not small, and their width effects may be important for the square of loop integral, which can be included by replacing $m_{i(j)}$ in Eq. (\ref{eq:1}) with $(m_{i(j)}-i\Gamma_{i(j)}/2)$.  In the fitting procedure, the schemes of fit I and fit II are adopted, which, as can be seen from Table \ref{table:parameter}, correspond to the same and independent cutoff parameters $\alpha$ for different combinations of intermediate charmonium states, respectively. It can be seen that the line shape of the LHCb data can be well described both in fit I and fit II, but the scheme of fit II locally performs better on peak structures. Anyway, in our fitting, three obvious peak structures near 6.5, 6.9, 7.3 GeV can be reproduced, which directly correspond to three rescattering channels $\eta_c\chi_{c1}$, $\chi_{c0}\chi_{c1}$ and $\chi_{c0} \chi_{c1}^{\prime}$. As for the channel of $J/\psi J/\psi$, it provides an effect of threshold enhancement, which explains the line shape behavior of experimental data near the threshold of $m_{J/\psi J/\psi}$. Furthermore, the peak position of $J/\psi h_c$ at 6.64 GeV precisely point out a jumping point in LHCb data as shown in Fig. \ref{fig:Fitexp}, whose verification can be treated as an interesting experimental topic for LHCb and CMS in the future under more accumulated experimental data. It is obvious that the above evidence provides a strong support to the non-resonant nature of several visible structures in the measurements of LHCb.  Additionally, when focusing on individual contributions in theoretically describing the line shape of LHCb data, it is easily found that the direct contribution of double $J/\psi$ production is dominant for the whole invariant mass distribution, which is reasonable because of loop suppression of the rescattering mechanism.

After studying the new observation in the invariant mass spectrum of a $J/\psi$-pair based on our non-resonant theoretical framework, we want to suggest an accessible way to further test the role of dynamical rescattering processes in the production of a $J/\psi$-pair. Similar to the case of di-$J/\psi$, such kind of rescattering processes should exist in the di-$\Upsilon$ production. Applying the input of resonance parameters from PDG \cite{Tanabashi:2018oca} (we refer to the theoretical estimates of Ref. \cite{Wang:2018rjg} for the unknown widths), the maximum positions of the channels $\eta_b\chi_{bJ}$ together with $\Upsilon h_b$, and $\chi_{bJ}\chi_{bJ}$ with $J=0,1,2$ are found to be clustered in the short energy interval of 19.28 to 19.36 and 19.73 to 19.83 GeV, respectively, and the channel $\Upsilon \Upsilon$ precisely leads to a cusp peak at 19.0 GeV. This means that three prominent peak structures near 19.0, 19.3, 19.7 GeV should be observed in the invariant mass spectrum of $m_{\Upsilon \Upsilon}$, where a double $\Upsilon$ can be reconstructed by $\mu^+\mu^-\mu^+\mu^-$ similar to the search for double $J/\psi$ events. 

In fact, although many theoretical calculations for spectroscopy of fully-charm tetraquark states including $J^{PC}=0^{++}$, $1^{-+}$, and $2^{++}$ can explain the measured mass of $X(6900)$, there are obvious differences on the predictions for the mass spectrum of the corresponding partner of fully-bottom tetraquark states \cite{Wu:2016vtq,Wang:2019rdo,Jin:2020jfc,Liu:2019zuc,Chen:2016jxd,Wang:2018poa,Anwar:2017toa,Chen:2018cqz,Heupel:2012ua,Bedolla:2019zwg,Karliner:2016zzc,Karliner:2017qhf,Berezhnoy:2011xn,Lu:2020cns,Yang:2020rih,Deng:2020iqw,Bai:2016int,Esposito:2018cwh,Chen:2019dvd,Hughes:2017xie}. On the other hand, because the heavy quark symmetry is assured, the mass difference among different combinations of intermediate heavy quarkonium states are almost independent of the heavy quark flavor. Thus, the measurements of production of $\Upsilon \Upsilon$ in high energy proton-proton collisions may be available to test the rescattering contributions and identify the nature of $X(6900)$ and other underlying structures, which should provide a good chance for LHCb and CMS.

%\section{Summary}\label{sec4}

{\it Conclusion.}---The LHCb Collaboration brought us some surprising results on the measurements of the invariant mass distribution of di-$J/\psi$ production, where there exist three obvious peaks between 6.194 and 7.4 GeV \cite{Aaij:2020fnh}. Although the explanation of fully-charm tetraquark states  for these structures is straightforward, their origins should still be investigated carefully because the peak phenomenon can also be produced by some special dynamical effects in addition to the resonance.

In this letter, we have proposed a non-resonant dynamical mechanism to understand several new structures observed by LHCb. Our idea  is based on a reaction that different combinations of a double charmonium directly produced in high energy proton-proton collisions are transferred into final states $J/\psi J/\psi$, which has been found to produce an obvious cusp at the corresponding mass threshold of a double charmonium. By fitting the experimental data by a line shape in the invariant mass spectrum of a $J/\psi$-pair, three obvious peak structures near 6.5, 6.9, and 7.3 GeV are well reproduced, which naturally correspond to three rescattering channels $\eta_c\chi_{c1}$, $\chi_{c0}\chi_{c1}$, and $\chi_{c0} \chi_{c1}^{\prime}$. Furthermore, we have predicted the peak line shape in the invariant mass spectrum of a $\Upsilon$-pair resulting from the similar rescattering processes between a double bottomonium, where three peak structures  near 19.0, 19.3, and 19.7 GeV could be detected in experiments, which are related to the rescattering channels of $\Upsilon \Upsilon$, $\eta_b\chi_{bJ}$ together with  $\Upsilon h_b$, and $\chi_{bJ}\chi_{bJ}$ with $J=0,1,2$, respectively. These predictions should be helpful to indirectly confirm the nature of the newly observed structures by LHCb.

In fact, the double $J/\psi$ or $\Upsilon$ is still an ideal final state to search for exotic fully-heavy tetraquark structures. Because our present knowledge for inner structures of tetraquark states is quite poor, it is hard to precisely know their masses just by calculations from a variety of phenomenological models.  However, there is a unique advantage for a peak behavior by the rescattering mechanism, i.e., their peak positions are very clear. Thus, combined with the fact that the degrees of freedom in a general four-body system are very considerable which can result in a very rich mass spectrum,
it is worth expecting to search for possible $QQ\bar{Q}\bar{Q}$ structures in which the corresponding peak positions are different from threshold masses of rescattering channels. We believe that precise measurements of LHCb and CMS in the future should have a chance to discover the definite candidate of fully-heavy tetraquark states {that may not be explained by our rescattering mechanism}, which will open a new chapter for understanding the complicated non-perturbative behavior of  QCD.

\section*{ACKNOWLEDGEMENTS}

This work is partly supported by the China National Funds for Distinguished Young Scientists under Grant No. 11825503, the National Natural Science Foundation of China under Grant No. 11775050, and by the Fundamental Research Funds for the Central Universities.

\end{document}